\documentclass[reprint,amsmath,amssymb,aps,floatfix]{revtex4-2}
\usepackage{graphicx} 
\usepackage{dcolumn} 
\usepackage{bm} 
\usepackage{mhchem} 
\usepackage{textcomp} 
\usepackage{color} 
\usepackage{textcomp}
\usepackage{xr}
\usepackage{float} 
\usepackage[colorlinks=true,linkcolor=blue,citecolor=blue,urlcolor=blue]{hyperref} 
\begin{document}

\preprint{APS/123-QED}

\title{\mbox Roughness-controlled Tribocharging Governs Friction in Dry Glass Contacts}

\author{Liang Peng$^{1}$\footnotemark[1]{*}}
\author{Begüm Demirkurt$^{1}$}  
\author{Thibault Roch$^{1}$} 
\author{Albert M. Brouwer$^{2}$} 
\author{Bart Weber$^{1,3}$} 
\author{Daniel Bonn$^{1}$} 
\affiliation{$^{1}$ Van der Waals-Zeeman Institute, Institute of Physics, University of Amsterdam, Science Park 904, 1098 XH Amsterdam, The Netherlands \\
$^{2}$ Van't Hoff Institute for Molecular Sciences, Science Park 904, 1098 XH Amsterdam, The Netherlands\\
$^{3}$ Advanced Research Center for Nanolithography (ARCNL), Science Park 106, 1098 XG Amsterdam, The Netherlands
}


\begin{abstract}
Friction is commonly reduced by polishing surfaces, based on the idea that roughness
enhances mechanical interlocking and thus friction. Here we show that, for dry glass–glass contacts, increasing nanoscale roughness can instead \emph{reduce} friction because it suppresses triboelectric adhesion. Using rheometer-based friction measurements in dry nitrogen, super-resolution imaging of the real contact area, soft x-ray discharge, and Faraday-cup electrometry, we demonstrate that sliding generates substantial tribocharges whose electrostatic attraction contributes significantly to friction. As the root-mean-square surface slope $h'_{\mathrm{rms}}$ of the glass ball is increased from 0.01 to 0.09, the real contact area and retained tribocharge both decrease strongly, while the average contact pressure increases by a factor of three, yet the friction coefficient drops by about 30\%. Discharging the interface with soft x rays largely removes the roughness dependence of friction. Our results show that nanoscale roughness controls tribocharging and electroadhesion in dielectric contacts, inverting the classical roughness–friction relation and identifying triboelectric effects as a key design parameter for friction control.
\end{abstract}

\maketitle

Friction governs the efficiency, stability, and lifetime of mechanical systems across length scales, from microelectromechanical systems and precision positioning devices to granular flows and large-scale industrial machinery. A central paradigm in tribology is that roughness enhances mechanical interlocking between asperities, so that smoother surfaces generally exhibit lower friction. This picture is challenged for polished interfaces with nanoscale roughness, where adhesive interactions such as van der Waals forces \cite{delrio2005}, water capillary bridges \cite{peng2022nonmonotonic}, and covalent interfacial bonding \cite{peng2023controlling} can dramatically increase friction. Recent work has shown that in humid environments, increasing nanoscale roughness can reduce friction by suppressing capillary adhesion, leading to the counterintuitive result that rougher surfaces can be more slippery \cite{hsia2021rougher}. An important open question is whether a similar inversion of the roughness–friction relation can arise in \emph{dry}, insulating contacts, where triboelectric charge generation \cite{jimidar2025enduring,LacksSankaran2011,sobolev2022charge,xu2024adhesion} and electrostatic adhesion are expected to play a central role.

Tribocharging \cite{pertl2025no,sobarzo2024multiple,sobarzo2025spontaneous,grosjean2023single,zhang2015electric,xu2024triboelectrification} is not merely a curiosity. It plays roles in granular flows \cite{lee2015direct}, planet formation \cite{steinpilz2020electrical}, triboelectric nanogenerators\cite{hu2024review}, and industrial powder handling\cite{huang2023modeling}. In sliding interfaces, the generated charge can in principle modify adhesion and friction through electrostatic forces\cite{peng2025polaritydependentelectroadhesionsiliconinterfaces}. However, direct experimental evidence linking tribocharging \cite{lacks2019long,Galembeck2014Review} to macroscopic friction—especially for nominally identical dielectric materials—remains limited.

Surface roughness plays a central role in multi-asperity friction, as it governs the real area of contact and the distribution of interfacial gaps across length scales. Recent advances in contact visualization \cite{Demirkurt2024} and multiscale roughness characterization \cite{pradhan2025surface} have enabled quantitative comparison between measured contact areas and contact mechanics theories, highlighting the importance of small-scale surface slopes and elastic deformation in determining the true contact fraction\cite{persson2001elastoplastic,terwisscha2024elastic}. In glassy and oxide systems under ambient conditions, small interfacial gaps can additionally give rise to capillary adhesion, which enhances friction when the roughness amplitude becomes comparable to the range of capillary forces\cite{hsia2021rougher,peng2022nonmonotonic}.

However, for interfaces whose roughness exceeds the nanometric range relevant for capillary condensation, capillary adhesion is expected to play a minor role. In such cases, the load-controlled friction coefficient\cite{berman1998amontons,hsia2021rougher} is expected to become independent of surface roughness as the interfacial adhesion is negligible compared to the external load. Whether this expectation holds for dielectric–dielectric interfaces such as glass–glass contacts remains an open question, particularly given the growing evidence that contact electrification and trapped charge can generate long-range electrostatic interactions at solid–solid interfaces\cite{musa2018charging,panda2022contact,sayfidinov2018minimizing,lee2018collisional}.

In this letter, we address this open question using glass–glass interfaces. By systematically varying and quantifying surface topography using atomic force microscopy (AFM), we isolate the role of surface roughness in both friction and charge generation. Super-resolution contact imaging provides direct insight into the distribution of real contact regions, while soft X-ray irradiation and Faraday cup measurements enable independent quantification of tribocharging during sliding. We find that smoother interfaces generate and retain significantly more charge and exhibit higher friction, whereas rougher interfaces are more slippery. Our results demonstrate that surface roughness governs the amount of tribocharging during sliding and establish electrostatic forces as a significant contribution to friction in dielectric interfaces.

\begin{figure*}[ht]
\includegraphics[width=0.75\textwidth]{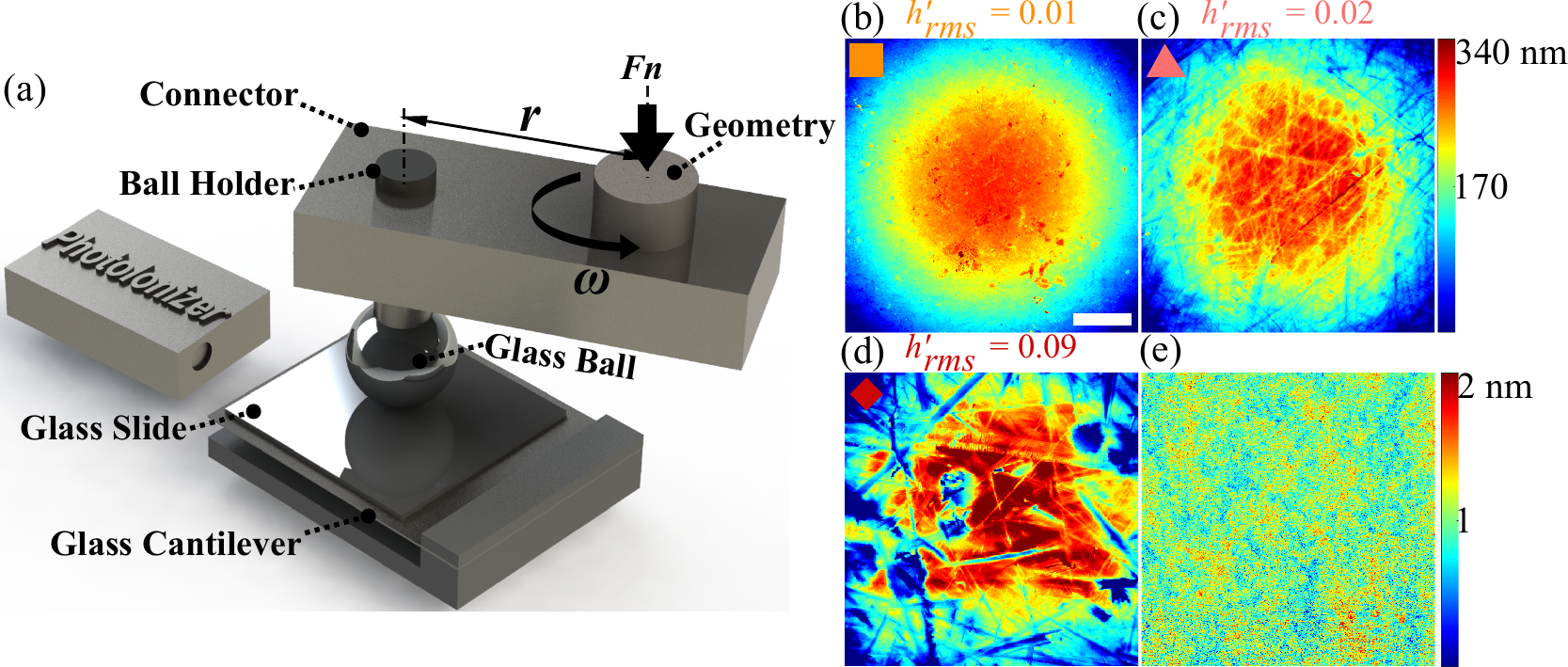}
\centering
\caption{ 
Experimental system and AFM topography. (a) Schematic of the rheometer-based glass–glass friction setup used in this study. A glass ball is slid against a smooth glass cantilever while the normal force ($F_n$) and friction force ($F_f$) are recorded simultaneously; a photoionizer is directed at the interface to neutralize sliding-induced tribocharges. (b–d) AFM tapping-mode topography measurements conducted on the glass balls that were roughened to varying degrees and (e) the smooth glass slide used in the experiments. The root-mean-square surface slope of the glass balls, $h'_{\mathrm{rms}}$, is 0.01, 0.02, and 0.09 respectively (also see the corresponding scan-size-dependent RMS roughness in Fig. S1). Scale bar, 10 $\mu$m.
} 
\label{fig1}
\centering
\end{figure*}

We measured friction between glass surfaces using a rheometer (DSR 502, Anton Paar), as shown in Fig.~\ref{fig1}(a) inside a dry (RH=0.8\%) nitrogen-purged chamber. A 3.68-mm-diameter soda-lime glass ball (Sigmund Lindner), clamped inside a ball holder, was mounted onto the rheometer geometry through a connector with an off-center distance $r=10$ mm and brought into contact with a homemade glass cantilever made by gluing a $25\times25$ mm microscope slide (VWR) to a steel sheet. The externally applied normal load $F_n$ was varied between 0--160~mN. During sliding, both normal force $F_n$ and dynamic friction force $F_f$ at the glass-glass interface were simultaneously recorded by the rheometer, from which the friction coefficient was obtained as $\mu=F_f/F_n$. The rheometer geometry was rotated at constant angular velocity $\omega$ to impose a sliding speed of $\omega r=0.25$ \textmu m/s at the contact interface; stable friction coefficients were reported after a few short sliding strokes of 5 \textmu m. In the friction measurements, glass balls with distinct nanoscale roughness were slid against much smoother glass microscopy slides (Fig.~\ref{fig1} and S1). The surface topography of the glass balls and the glass slide was measured by AFM (Dimension Icon, Bruker) in tapping mode using Si tips (RTESPA-300, Bruker) over an area of $50\times50$~\textmu m\textsuperscript{2}. The pixel sizes were 48.83 nm for the two smoother glass balls and 87.89 nm for the roughest glass ball, respectively. The data was analyzed to obtain the root-mean-square surface slope, $h'_\mathrm{rms}$. The three distinct topographies resulted in $h'_\mathrm{rms}$ values of 0.01, 0.02, and 0.09 as shown in Fig.~\ref{fig1}(b)--1(d). To remove tribocharges generated during sliding, the surfaces were neutralized using soft X-ray emission from a photoionizer (L12645, Hamamatsu). Real contact areas were independently measured at fixed load $F_n=100$ mN using super-resolution fluorescence microscopy, as detailed in Ref.~\cite{Demirkurt2024}. The spatial resolution of the super-resolution images, determined by Fourier ring correlation (FRC), is approximately $\sim$30 nm (see more details in Fig.~S2).

\begin{figure}[h]
\includegraphics[width=0.45\textwidth]{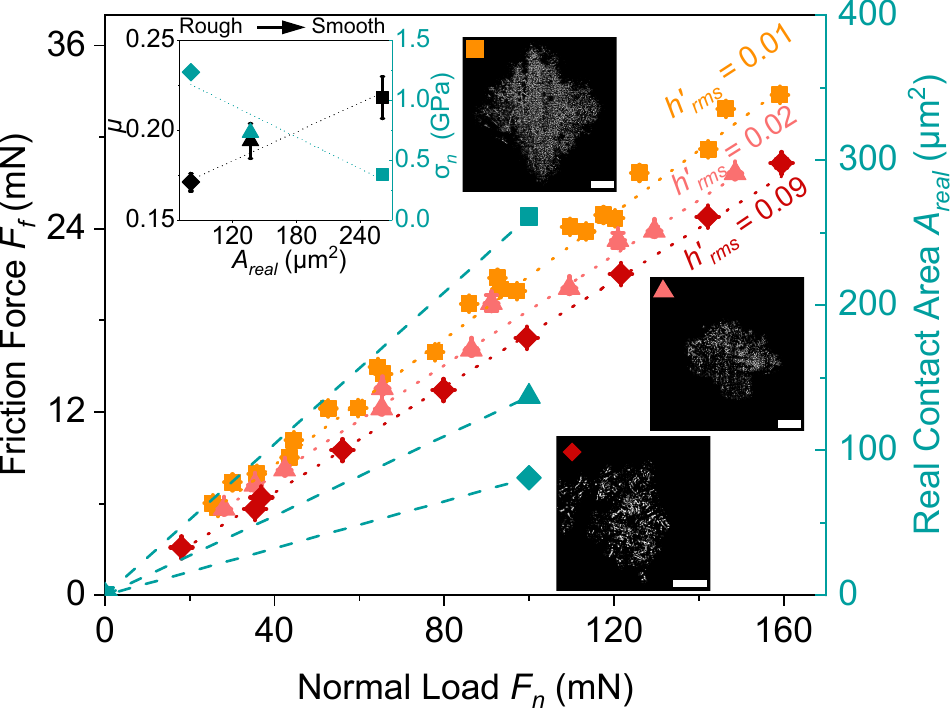}
\centering
\caption{ 
Friction and real contact area measurement. Friction force, $F_f$, was measured as a function of externally applied normal load, $F_n$, in a dry nitrogen environment (RH=0.8\%). Square, triangle and diamond symbols represent measurements obtained using glass balls with $h_{rms}$ of 0.01, 0.02 and 0.09, respectively. The corresponding real contact areas, measured using super-resolution fluorescence imaging with a resolution of \(\sim 30\) nm, under an applied load of 100 mN are shown as insets with corresponding symbols. In the real contact area maps, the white and black area indicate the contact and noncontact area, respectively. The top-left inset illustrates the variation of friction coefficient and contact pressure as a function of the measured real contact area. The contact pressure is calculated as the ratio between the applied load, 100 mN, and the real contact area. All scale bars represent 10 \textmu m.
} 
\label{fig2}
\centering
\end{figure}

Figure~\ref{fig2} summarizes the load dependence of friction and the corresponding real contact area for glass–glass contacts with three different nanoscale roughnesses. For all three balls, the friction force $F_f$ increases linearly with externally applied normal load $F_n$ over the entire range of loads explored, indicating Amontons'-like behavior despite the significant differences in surface topography (Fig.~\ref{fig1}(b)--1(d)). Super-resolution fluorescence imaging at $F_n=100$ mN with $\sim$30 nm lateral resolution at glass-glass contacts reveal that the real contact area $A_\mathrm{real}$ decreases strongly with increasing roughness: the smoothest ball forms an extended, nearly contiguous contact, whereas the roughest ball only touches the glass slide in a sparse population of small asperity contacts (Fig.~\ref{fig2}). The resulting average contact pressure $\sigma_n=F_n/A_\mathrm{real}$ spans approximately a factor of three across the three roughnesses, while the friction coefficient $\mu=F_f/F_n$ changes by only a factor $\simeq1.4$ (top-left inset in Fig.~\ref{fig2}). This weak dependence of $\mu$ on the area of real contact suggests that the friction is load controlled\cite{hsia2021rougher}.

A key observation is that the friction coefficient decreases with increasing $h'_\mathrm{rms}$, demonstrating that rougher surfaces can be more slippery. This reduction is consistent with a suppression of adhesive interactions driven by a roughness induced increase in the separations or gaps between the sliding surfaces\cite{hsia2021rougher,peng2022nonmonotonic}. Given that the measurements are performed in dry nitrogen (RH=0.8\%) and in the absence of lubricants or capillary bridges\cite{peng2023controlling}, van der Waals interactions and triboelectric charges are the most likely sources of adhesion. In contrast to van der Waals or capillary adhesion, which decay rapidly with separation, electrostatic adhesion between surfaces carrying a fixed charge density can remain substantial even across larger gaps. In the simplest parallel-plate model, electroadhesion pressure scales with the square of the charge density and is independent of the separation distance\cite{persson2021general}. 

\begin{figure}[h]
\includegraphics[width=0.4\textwidth]{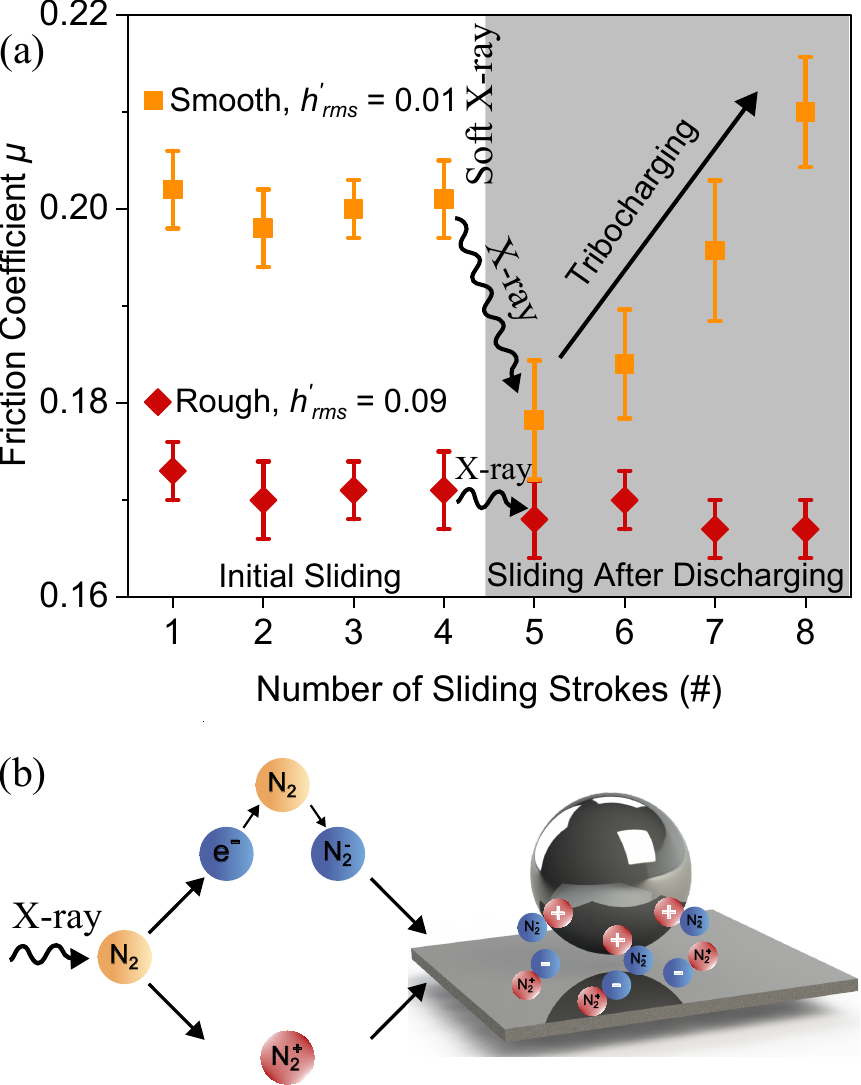}
\centering
\caption{ 
Soft X-ray mediated friction variation. (a) Evolution of friction coefficient, \(\mu\), over sliding strokes for smooth (yellow squares) and rough (red diamonds) glass–glass contacts in dry nitrogen under an externally
applied load of 20 mN. After four sliding strokes (5 \textmu m each), the contact is separated and exposed to soft X-ray irradiation from a photoionizer before sliding is resumed at the same contact location, as indicated by the shaded region (see more measurements in Fig.~S2).
(b) Schematic of the mechanism by which soft X-ray irradiation discharges the surfaces.
} 
\label{fig3}
\centering
\end{figure}

To directly probe the role of tribocharging in friction, we monitored the evolution of the friction coefficient over successive sliding strokes and used soft X-ray irradiation to  discharge the interface (Fig.~\ref{fig3}) between strokes for the smoothest and the roughest contacts. At the initial four sliding strokes, the rough contact shows a lower \(\mu\) than the smooth contact, consistent with the results in Fig.~\ref{fig2}. However, after discharging the separated interface using soft X-ray irradiation, $\mu$ for the smooth contact drops dramatically, whereas it is less affected for the rough contact (Fig.~\ref{fig3}(a)). 

During discharging, soft X-rays ionize the surrounding \(N_2\) molecules, producing both positive \(N^+_2\) and negative \(N^-_2\) ions. These ions neutralize the electrostatic charges accumulated on the glass ball and substrate during the initial sliding process (Fig.~\ref{fig3}(b)). As a result, the electrostatic adhesion is suppressed. This discharge–recharge behavior provides strong evidence that sliding-induced tribocharging contributes substantially to the measured friction. Notably, after discharging, the friction coefficients of the smooth and rough contacts become nearly identical. Such behavior is consistent with friction enhancement arising from electrostatic attraction between tribocharged surfaces, rather than from van der Waals interactions, indicating a history-dependent electrostatic contribution to friction.

\begin{figure}[h]
\includegraphics[width=0.4\textwidth]{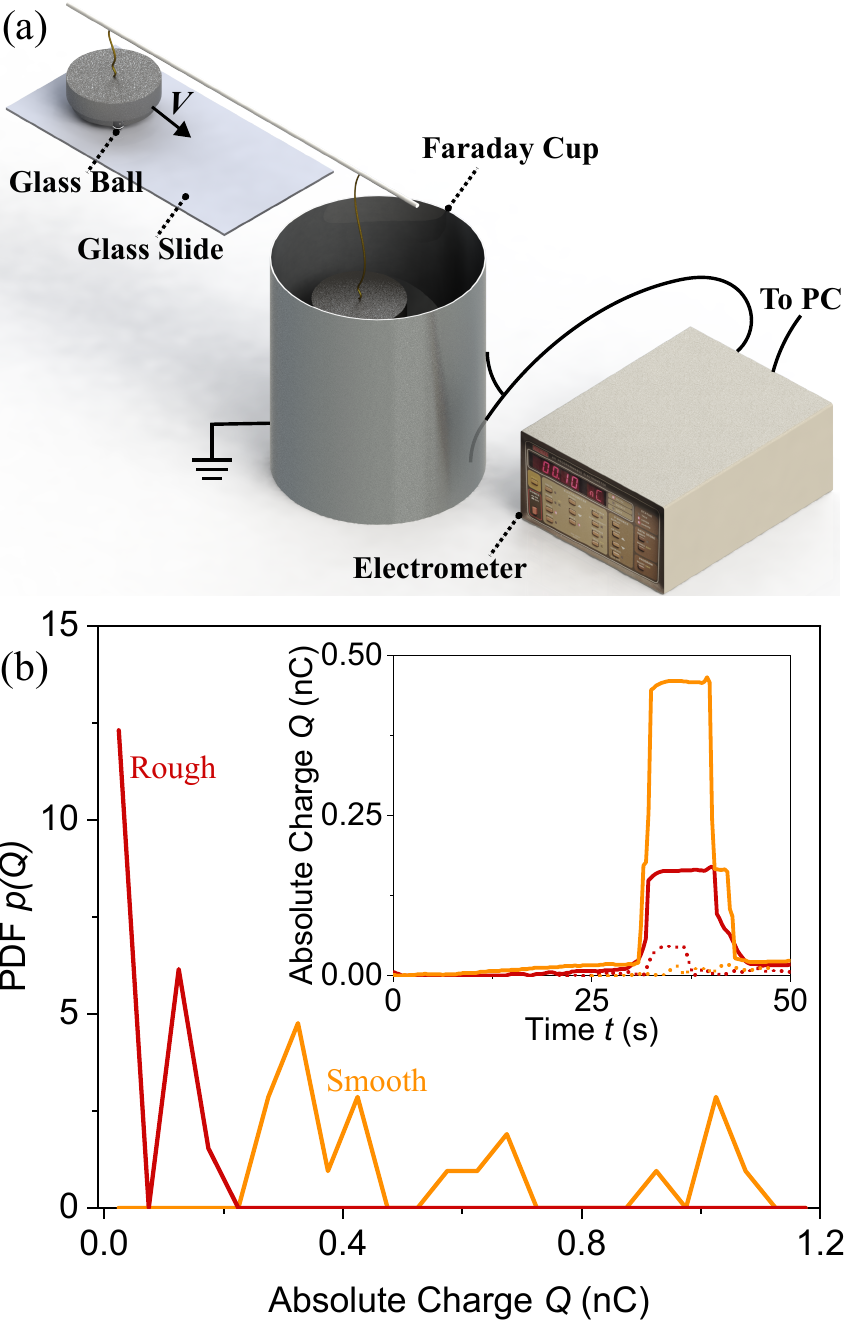}
\centering
\caption{Tribocharge measurement. (a) Experimental setup. Three discharged glass balls, clamped within a holder, were slid 20 mm against a discharged glass slide at a load of 415 mN before the slider was manually transferred into a Faraday cup. The holder was suspended inside the cup until measured charge stabilized and was then withdrawn. The outside conducting layer of the Faraday cup was grounded and was connected to the inner conducting layer of the cup through a Keithley 617 electrometer to measure the accumulated tribocharge on the balls. The entire setup was placed inside a dry chamber (RH=0.8\%) and a computer connected to the electrometer was used to simultaneously log the charge. (b) The probability density function, \textit{p(Q)}, of the absolute tribocharge, \textit{Q}, at the smooth (yellow) and rough (red) contacts. The inset shows examples of the real time evolution of measured charge during insertion of the slider into the Faraday cup, using the same color scheme, with solid and dashed curves representing the measurements with and without sliding before insertion into the Faraday cup, respectively. 
} 
\label{fig4}
\centering
\end{figure}

To rationalize the contrast in friction drop between rough and smooth interfaces upon x-ray irradiation, we performed independent tribocharge measurements using a Faraday cup electrometer (Fig.~\ref{fig4}(a)). In these experiments, three discharged glass balls were slid over a discharged glass slide inside a dry chamber. The sliding distance was 20 mm and the sliding speed was 4 mm/s. A total normal load of 415 mN was applied during sliding. After the measurement, the ball–holder assembly was rapidly transferred into a grounded Faraday cup connected to a sensitive electrometer (Keithley 617). The time-resolved charge signal was recorded until it reached a steady value before the assembly was withdrawn from the cup. The absolute tribocharge $Q$ was then extracted as the difference between the instantaneous charge and the baseline. Repeating this protocol for many nominally identical runs yielded the probability density function $p(Q)$ of $Q$ for both smooth and rough contacts (Fig.~\ref{fig4}(b)). The distributions show that sliding generates significantly larger absolute charges on the smooth ball than on the rough one: the $p(Q)$ of the smooth contact is broader and shifted towards higher $Q$, whereas the rough contact exhibits a narrower distribution concentrated at smaller $Q$ values. Control experiments without prior sliding (dashed curves in the inset of Fig.~\ref{fig4}(b)) confirm that the measured charges indeed arise from triboelectrification rather than from handling or environmental contamination. 

The substantially larger tribocharge measured for the smooth contact demonstrates that smoother interfaces are capable of generating and/or retaining more triboelectric charge than rougher interfaces. A natural first hypothesis is that this difference originates primarily from the larger real area of contact. Since tribocharging is expected to occur locally within the real contact regions\cite{waitukaitis2026}, smoother interfaces with larger contact areas should indeed charge more rapidly during sliding. However, the friction evolution in Fig.~\ref{fig3} suggests that the smooth contact approaches a steady tribocharged state already after a few sliding strokes of only 5~\textmu m. The rough contact possesses a real contact area that is smaller by approximately a factor of three (Fig.~\ref{fig2}), and may therefore charge proportionally more slowly. Nevertheless, the sliding distance used in the Faraday cup experiments (\(20~\mathrm{mm}\)) is several orders of magnitude larger than the sliding distance required to establish the friction contrast in Fig.~\ref{fig3}. This suggests that differences in charging rate alone are unlikely to explain the strong contrast in retained tribocharge observed between smooth and rough interfaces.

Instead, the dominant difference may arise from how efficiently tribocharge is retained at the interface. Rough interfaces contain larger interfacial gaps and steeper local asperities than smooth interfaces. For a given transferred charge density, these larger separations produce larger local potential differences across the interface, potentially enabling field-assisted discharge, gas ionization, or dielectric breakdown processes that neutralize the tribocharge~\cite{zhang2015electric,sobolev2022charge}. In addition, the larger local surface slopes and sharper asperities present at rough interfaces may locally enhance the electric field, thereby facilitating charge emission or discharge processes analogous to field-enhanced dielectric breakdown. Smooth interfaces, which contain smaller gaps and lower local slopes~\cite{terwisscha2024elastic}, may therefore sustain electrostatically coupled charge distributions more effectively during sliding. In this picture, rough interfaces not only generate less tribocharge because of their smaller real contact area, but also lose charge more efficiently through discharge processes across the larger interfacial gaps and at high-field asperities. This provides a natural explanation for both the substantially smaller retained tribocharge measured in Fig.~\ref{fig4} and the weaker friction enhancement observed for rough contacts.

The relation between the measured tribocharge and the resulting electroadhesion is, however, not straightforward. A simple estimate illustrates that the measured net tribocharge cannot remain fully electrostatically coupled across the sliding interface. If the typical tribocharge measured for the smooth contact (Fig.~\ref{fig4}, \(\sim 0.6~\mathrm{nC}\)) were assumed to reside as uniformly opposed charge distributions within the Hertzian contact area (3012 \textmu m\textsuperscript{2}), a parallel-plate estimate predicts an electroadhesion force of 6807 mN, sufficiently large to strongly hinder sliding (see more details in the Supplemental Material, Sec. A2). This is inconsistent with the experimentally observed steady sliding behavior in the Faraday cup experiments (Fig.~\ref{fig4}). This observation implies that only a fraction of the measured tribocharge contributes directly to the interfacial electrostatic attraction. The remaining charge is likely redistributed over regions that are electrostatically weakly coupled to the opposing surface, for example through lateral charge migration\cite{NAVARRORODRIGUEZ2023155437} away from the contact region. Since the Faraday cup measurements only provide the total retained charge after separation, they do not reveal the spatial distribution of charge during sliding. Spatially resolved measurements of interfacial charge distributions will therefore be essential for establishing a quantitative connection between tribocharging, electroadhesion, and friction\cite{peng2025polaritydependentelectroadhesionsiliconinterfaces}.


In conclusion, we have shown that nanoscale roughness strongly influences triboelectric 
friction in dry glass–glass contacts. Smooth interfaces exhibit larger real contact areas, retain substantially more tribocharge, and display stronger friction enhancement than rough interfaces. Increasing the root-mean-square surface slope reduces both the real contact area and the amount of tribocharge that remains electrostatically coupled across the interface, while simultaneously promoting discharge across larger gaps and at high-field asperities. As a result, the electrostatic contribution to friction is suppressed and rougher interfaces become more slippery, even though the average contact pressure is higher. These findings establish tribocharging and electrostatic adhesion as important contributors to friction in dielectric contacts, and show that nanoscale roughness can invert the classical roughness–friction relation by frustrating electroadhesion. At the same time, a simple parallel-plate estimate using the measured net charge demonstrates that only a fraction of the retained charge can remain coupled across the interface, underscoring the need for spatially resolved measurements of interfacial charge distributions to quantitatively link tribocharging, electroadhesion, and friction. More broadly, our results call for microscopic theories of friction in insulating materials that explicitly incorporate the coupled evolution of real contact area, charge generation, and charge retention.

\textit{Acknowledgments}---This work was supported by the European Research Council (ERC) under the European Union’s Horizon 2020 research and innovation program (Grant No. 833240). This research was funded in part by the Swiss National Science Foundation (SNFS) P500PT/222342. For the purpose of Open Access, a CC BY public copyright license is applied to any Author Accepted Manuscript (AAM) version arising from this submission

\bibliography{Reference}

\end{document}


\preprint{APS/123-QED}

\title{\mbox Supplemental Material for: Roughness-controlled Tribocharging Governs Friction in Dry Glass Contacts}

\author{Liang Peng$^{1}$\footnotemark[1]{*}}
\author{Begüm Demirkurt$^{1}$}  
\author{Thibault Roch$^{1}$} 
\author{Albert M. Brouwer$^{2}$} 
\author{Bart Weber$^{1,3}$} 
\author{Daniel Bonn$^{1}$} 
\affiliation{$^{1}$ Van der Waals-Zeeman Institute, Institute of Physics, University of Amsterdam, Science Park 904, 1098 XH Amsterdam, The Netherlands \\
$^{2}$ Van't Hoff Institute for Molecular Sciences, Science Park 904, 1098 XH Amsterdam, The Netherlands\\
$^{3}$ Advanced Research Center for Nanolithography (ARCNL), Science Park 106, 1098 XG Amsterdam, The Netherlands
}

\maketitle
\onecolumngrid 

\subsection*{Section A: Experimental Methods}

\vspace{1em}
\noindent\textbf{A1. Fluorescence Imaging of Contact Area}

 The contact was established by pressing the glass ball used in the friction measurements against a microscopy glass coverslip coated with a DCDHF dye monolayer at a normal load of 100~mN, using an Anton Paar DSR 301 rheometer, as detailed in our previous study~\cite{demirkurt2024resolving}. An inverted, home-built, open-frame wide-field/TIR epifluorescence microscopy setup was used to image the contact interface. A Coherent Sapphire LP488 laser operating at 488~nm was used for illumination. In all measurements, the power density was kept constant at 1~kW~cm\textsuperscript{-2}. A 535/50~nm emission filter and a 488~nm notch filter were used in the detection channel. The objective was an Olympus UAPON 100$\times$ oil-immersion objective with NA~1.49. The emitted fluorescence was recorded using a Hamamatsu C11440s CMOS camera at an acquisition rate of 100~ms per frame.

After the contact was established and a diffraction-limited fluorescence image was recorded, the DCDHF monolayer in the contact region was photobleached under load using intense continuous-wave 488 nm illumination until sparse single-molecule blinking was observed, allowing super-resolution imaging. Individual blinking events were then recorded over approximately 20,000 frames at an acquisition rate of 100~ms per frame. The positions of the blinking molecules were localized by fitting their point-spread functions, and the accumulated molecular coordinates were used to reconstruct the super-resolution contact image. The reconstruction was performed by rendering each localization as a two-dimensional Gaussian function, with the standard deviation set by the final image resolution determined from Fourier ring correlation analysis, as shown in Fig.~S3.

\vspace{1em}
\noindent\textbf{A2. Hertzian contact area and electroadhesion calculation}

For the three-ball-on-glass slide contact under a total load of 415 mN, the total Hertzian contact area, \(A_{Hertz}^{total}\), is calculated as:

\begin{equation}
\begin{aligned}
{A_{Hertz}^{total} =3\times A_{Hertz}=3\times\pi\times(\frac{3RF_n}{4E^*} )^{2/3}} 
\label{eq4}
\end{aligned}
\end{equation}
where \({E^*}\) is the the composite elastic modulus and defined as \(\frac{1}{E^*} = \frac{1-\nu_1^2}{E_1}+\frac{1-\nu_2^2}{E_2}\). \({E_1}, {\nu_1}\) and \(E_2, \nu_2 \) are the Young’s modulus and Poisson’s ratio of the two contacting bodies. For the glass-glass contact here, \(E_1=E_2\)=64 GPa and \(\nu_1=\nu_2\)=0.2. \(R\) and \(F_n\) are the radius of the glass ball and the load carried by each ball, respectively, with \(R=1.84\) mm and \(F_n=415/3 \approx 138\) mN. The resulting total Hertzian contact area is \(A_{Hertz}^{total}=3\times1004\) \textmu m\textsuperscript{2}= 3012 \textmu m\textsuperscript{2}.

To estimate the electroadhesion force generated in the charge measurements (Fig.~4), we approximate the charged interface as a parallel-plate capacitor, taking the total macroscopic Hertzian contact area (\(A_{Hertz}^{total}=3012\) \textmu m\textsuperscript{2}) as the capacitor area. This parallel-plate approximation is justified by assuming a uniform tribocharge density, in which the electroadhesion is independent of the local interfacial separation distance caused by the roughness\cite{persson2021general}. The electroadhesion force, \(F_{ea}\), can thereby be calculated: 
\begin{equation}
\begin{aligned}
{
F_{ea}=\frac{1}{2}\epsilon_0\kappa E^2A_{Hertz}^{total}=\frac{1}{2}\epsilon_0\kappa(\frac{\sigma}{\epsilon_0\kappa})^2A_{Hertz}^{total}
}
\label{eq4}
\end{aligned}
\end{equation}
where \(\epsilon_0\) and \(\sigma=\sim 0.6\) nC are the vacuum permittivity and charge density, respectively; \(\kappa=1\) is elative permittivity of air. This yields the resulting electroadhesion force of 6807 mN.

\newpage
\subsection*{Section B: Experimental Results}
\renewcommand{\thefigure}{S\arabic{figure}}

\vspace{1em}

\begin{figure*}[th]
\includegraphics[width=0.5\textwidth]{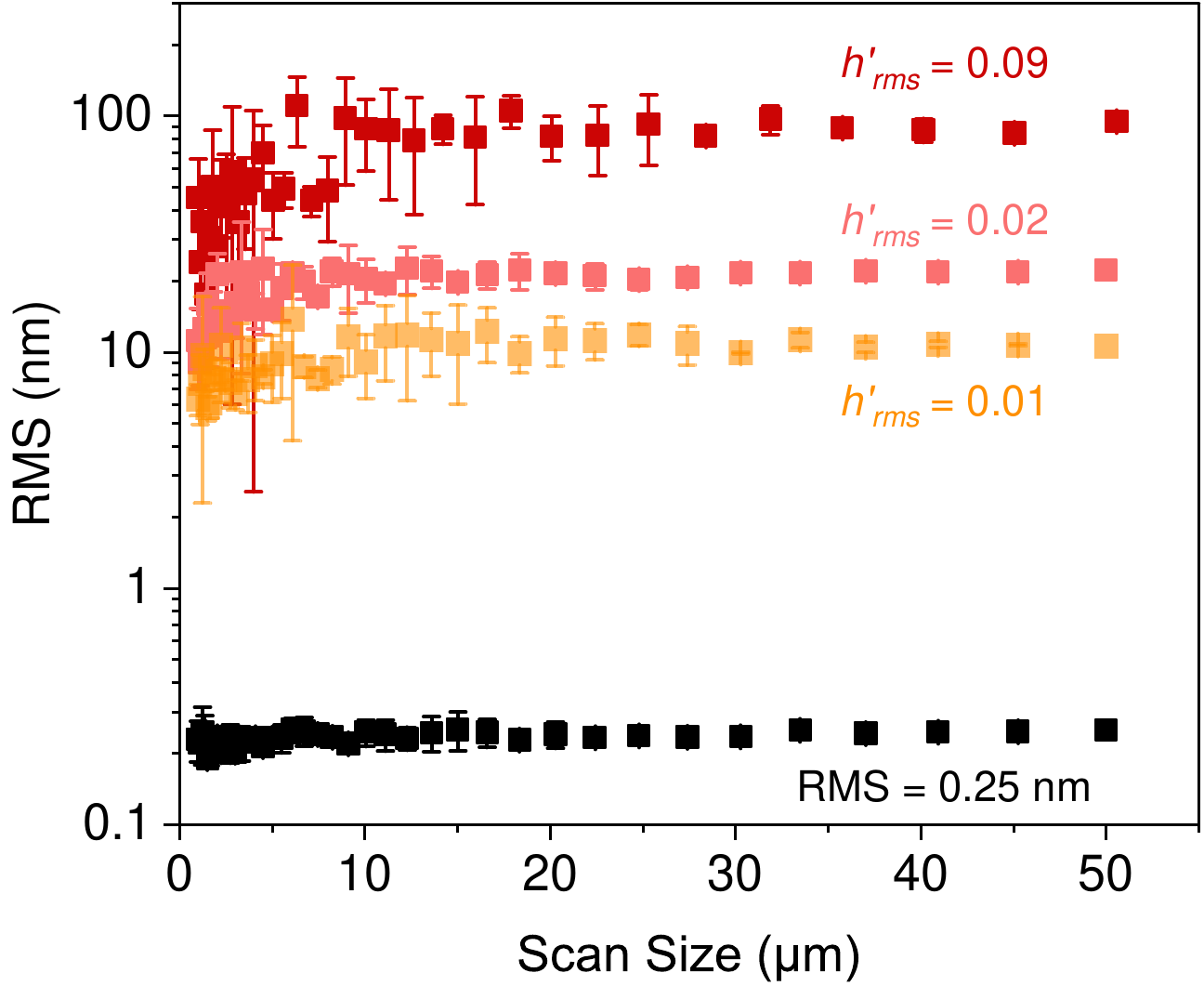}
\centering
\caption{ 
Scan size-dependent RMS roughness. The RMS roughness is plotted as a function of scan size for the topography of the three rough glass balls and the glass slide. The full scan size is $50\times50$~\textmu m\textsuperscript{2}. Each data point represents the average value obtained from three randomly selected subareas cropped from the full AFM topography. Red, pink and yellow correspond to the roughest, moderately rough and smooth glass balls; Black represents the glass slide.
} 
\label{figs1}
\centering
\end{figure*}

\begin{figure*}[th]
\includegraphics[width=0.5\textwidth]{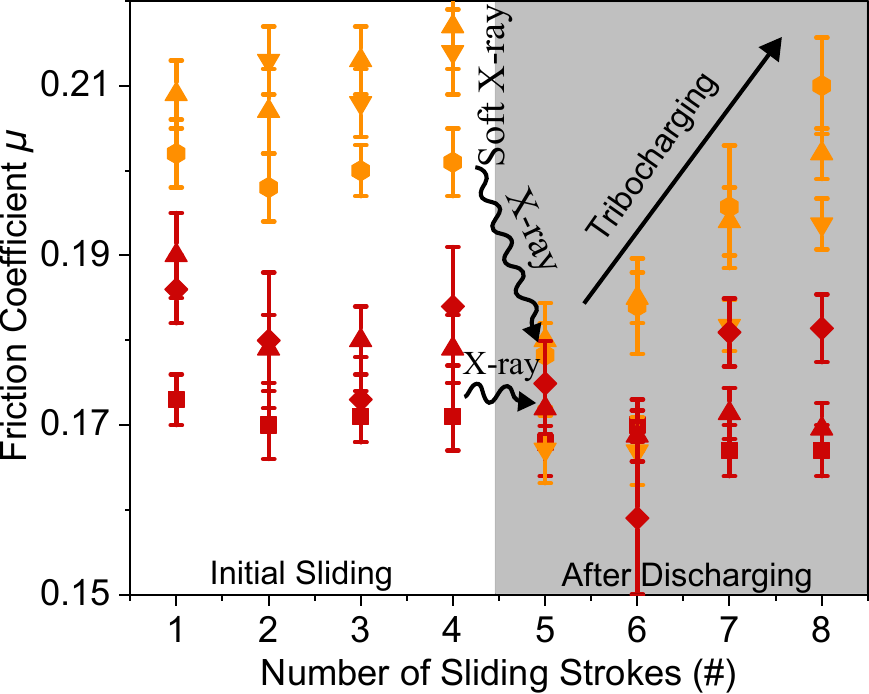}
\centering
\caption{ 
Evolution of friction coefficient, \(\mu\), over sliding strokes for smooth (yellow) and rough (red ) glass–glass contacts in dry nitrogen under an externally applied load of 20 mN. Different symbols represent different measurements.  
} 
\label{figs2}
\centering
\end{figure*}

\begin{figure*}[th]
\includegraphics[width=0.9\textwidth]{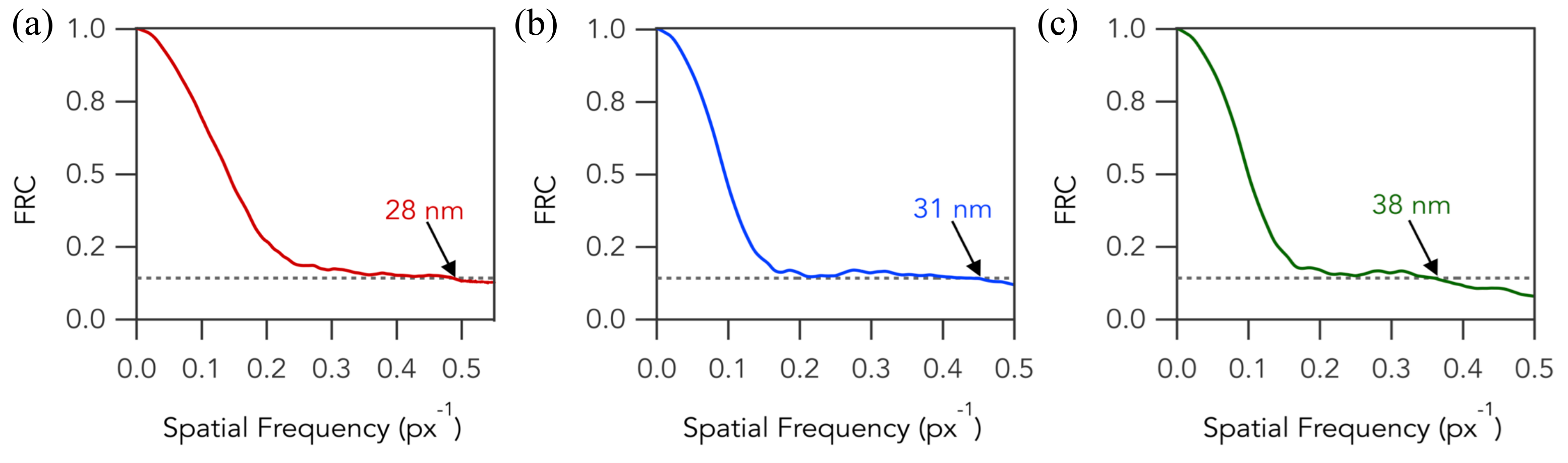}
\centering
\caption{ 
FRC map of the super-resolution images. (a), (b), and (c) correspond to the three rough contacts with surface root-mean-square slope $h'_{{rms}}=$  0.01, 0.02, and 0.09, respectively.
} 
\label{figs3}
\centering
\end{figure*}

\clearpage
\newpage

\bibliography{Reference}